\documentclass{article}
\usepackage{epsfig}
\begin{document}
\title{\bf  Density matrix for a consistent non-extensive thermodynamics}
\author{Marcelo R. Ubriaco\thanks{Electronic address:ubriaco@ltp.uprrp.edu}}
\date{Laboratory of Theoretical Physics\\Department of Physics\\University of Puerto Rico\\17 Ave. Universidad Ste. 1701\\
San Juan, PR 00925-2537, USA}

\maketitle
\begin{abstract}
Starting with the average particle distribution function for bosons and fermions for non-extensive thermodynamics , as proposed in \cite{CMP}, we obtain the corresponding density matrix operators 
and hamiltonians.  In particular, for the bosonic case the corresponding operators satisfy a deformed bosonic algebra and the hamiltonian involves interacting terms in powers of  $a^{\dagger}_ja_j$ standard creation and annihilation operators.
For the unnormalized density matrix we obtain a nonlinear equation that leads to a two-parameter solution relevant to anomalous diffusion phenomena.

\end{abstract}
{\it Keywords:}\\
 Non-extensive thermodynamics\\
Density matrix\\
Entropy functions\\
Anomalous diffusion.

\section{Introduction}
Since its formulation, non-extensive thermodynamics \cite{T}  has found applications to a vast number of fields \cite{T1}. Recently, some applications include
 proton-proton collisions\cite{CW} and neutron stars \cite{MMD1} \cite{MDMC}.  These applications  are based on the claim of thermodynamic  consistency  leading to a modified fermionic  particle number distribution \cite{CMP} which consists of the $q$-power of the original Tsallis distribution.  In this manuscript, we obtain a factorized density matrix and   partition function for the bosonic and fermionic cases. Then, we show that the unnormalized density matrix leads to a generalized nonlinear  differential equation relevant to anomalous diffusion.
In Section \ref{DM} we obtain the density matrices and partition functions in a factorized form for boson and fermion cases. In particular, for the boson case the adjoint operators satisfy a deformed algebra leading to an interacting  hamiltonian. In Section \ref{S} we find a thermodynamic expression for the bosonic and fermionic entropy functions, and in Section \ref{AD}
we obtain and solve  a two-parameter nonlinear differential equation relevant to anomalous diffusion. Section 5 contains our concluding remarks.

\section{Density Matrix}\label{DM}
In this Section we obtain the density operator  for boson and fermions according to the particle distribution function defined  in \cite {MMD1} and the corresponding partition functions.  Let us define the function
\begin{equation}
\rho_k=\left(1+(q-1)\beta\epsilon'_k\right)^{\frac{1}{q-1}},
\end{equation}
where $\epsilon_k'=\epsilon_k-\mu$.  We want to obtain,  by using the operator formalism, the  density operator $\hat{\rho}$ that will give the number distributions:
\begin{equation}
<N_k>=\frac{1}{(\rho_k\pm 1)^q},\label{n}
\end{equation}
where the upper (lower)   sign are for fermions (bosons), and as it is well known these functions  become the fermion and boson number  distribution in the $q\rightarrow 1$ limit.
From the definition
\begin{equation}
<N_k>=Tr\hat{\rho}\phi_k^{\dagger}\phi_k,
\end{equation}
where $\phi$ and its adjoint $\phi^{\dagger}$  denote either  boson or fermion operators and with use of the usual commutation and anticommutation relations we find that
\begin{equation}
<N_k>=\mp Tr \phi_k^{\dagger}\hat{\rho}\phi_k\pm 1,
\end{equation}
 Requiring that the density matrix $\hat{\rho}$ satisfy the  relations with the operator $\phi_k^{\dagger}$
\begin{equation}
\phi_k^{\dagger}\hat{\rho}=\mp \hat{\rho} \phi_k^{\dagger}+(\rho_k\pm 1)^q\hat{\rho}\phi_k^{\dagger}, \label{phirho}
\end{equation}
leads to 
\begin{equation}
<N_k>=\pm1+<N_k>\mp(\rho_k\pm 1)^q<N_k>,
\end{equation}
and
\begin{equation}
0=\pm1\mp(\rho_k\pm 1)^q <N_k>,
\end{equation}
and then to Eq. (\ref{n}).
\subsection{Boson case}
For the boson case, Eq. (\ref{phirho}) reads
\begin{equation}
a_k^{\dagger}\hat{\rho}=\left(1+(\rho_k-1)^q\right)\hat{\rho}a_k^{\dagger},
\end{equation}
such that considering the scaling operator $\Lambda^{\hat{n_k}}=\Lambda^{-1}a_k\Lambda^{\hat{n_k}}$, where $\hat{n_k}=a^{\dagger}a$ is the usual boson number operator, we see that Eq. (\ref{phirho}) is satisfied by the following operator
\begin{equation}
\hat{\rho}=\frac{1}{Z}\prod_{k=0}\left(1+(\rho_k-1)^q\right)^{-\hat{n}_k},\label{rho}
\end{equation}
where $Z$ is the partition function
\begin{equation}
Z=\sum_{n_1=0}...\sum_{n_{\infty}=0}\prod_{k=0}\left(1+(\rho_k-1)^q\right)^{-\hat{n}_k},
\end{equation}
leading to the product
\begin{equation}
Z=\prod_{k=0}\frac{1}{1-\left(1+(\rho_k-1)^q\right)^{-1}}.
\end{equation}
From Eq. (\ref{rho}) we can find the corresponding hamiltonian  $\hat{H}_k$  by equating 
\begin{equation}
\left(1+\beta (q-1)\hat{H}_k\right)^{\frac{1}{1-q}}=\left(1+(\rho_k-1)^q\right)^{-\hat{n}_k},\label{H}
\end{equation}
where the left hand side becomes an   exponential of the energy as $q\rightarrow 1$, giving after a simple manipulation
\begin{equation}
\hat{H}_k=\frac{\left(1+(\rho_k-1)^q\right)^{-(1-q)\hat{n}_k}-1}{\beta(q-1)},\label{Hk}
\end{equation}
which as a power series  the hamiltonian becomes
\begin{equation}
\hat{H}_k=\sum_{l=1}\frac{(a^{\dagger}_ka_k)^l}{\beta} (q-1)^{l-1}log^l(1+(\rho_k-1)^q) \label{Hk}
\end{equation}
indicating  that this model contains interaction terms involving an even number of operators.  A simple check shows that, as expected, 
$\hat{H}_k\rightarrow \hat{n}_k\epsilon'_k$ as $q\rightarrow 1$.

We can  rewrite Eq. (\ref{Hk}) for the full hamiltonian in a standard form in terms of two adjoint operators
\begin{equation}
\hat{H}=\sum_k \epsilon'_k\overline{\Phi}_k\Phi_k,
\end{equation}
if we  define 
\begin{equation}
\overline{\Phi}_k=a_k^{\dagger},\;\;\;
\Phi=a^{-1\dagger}_k\frac{Q_k^{\hat{n}_k}-1}{\rho_k^{q-1}-1},
\end{equation}
where  $Q_k=\left(1+(\rho_k-1)^q\right)^{q-1}$, leading to the deformed bosonic  algebra
\begin{equation}
\overline{\Phi}_j\Phi_k-Q^{-1}_k\Phi_k\overline{\Phi}_j=\delta_{j,k} \frac{Q^{-1}_k-1}{\rho^{q-1}_k-1}.
\end{equation}
The deformation of this boson algebra, in contrast to 
 the usual $q$-boson algebras \cite{LS}, depends on the value of the energy. In particular, for $\rho_k=1$ and $q\neq 1$ the operators commute.

\subsection{Fermion case}
In this case, Eq. ({\ref{phirho}) is satisfied by the density operator
\begin{equation}
\hat{\rho}=\frac{1}{Z}\prod_k \left((\rho_k+1)^q-1\right)^{-\hat{n}_k}, \label{rhoF}
\end{equation}
leading to
\begin{equation}
\hat{\rho}=\frac{1}{Z}\prod_k\left(1+\hat{n}_k(\Gamma^q_k-1)^{-1}-1\right),
\end{equation}
where $\Gamma_k=(1+\rho_k)$, $\hat{n}_k=b^{\dagger}_kb_k$ is the usual fermion number operator and the partition function
\begin{equation}
Z=\sum_{n_0=0}^{1}...\sum_{n_\infty=0}^{1}\prod_{k=0}\left(1+\hat{n}_k[(\Gamma^q_k-1)^{-1}-1]\right),
\end{equation}
 becomes 
\begin{equation}
Z=\prod_{k=0}\frac{1}{1-(1+\rho_k)^{-q}}
\end{equation}
In this case equating  the left hand side of Eq.(\ref{H}) with the corresponding factor of the product in Eq.(\ref{rhoF}),  we find that the hamiltonian is given by
\begin{equation}
\hat{H_k}=\frac{\left(\Gamma_k^q-1\right)^{(q-1)\hat{n_k}}-1}{\beta(q-1)},
\end{equation}
which reduces to the simple expression
\begin{equation}
\hat{H_k}=\frac{\hat{n_k}\left((\Gamma_k^q-1)^{q-1}-1\right)}{\beta(q-1)},
\end{equation}
implying that defining "new" fermionic operators will not lead to a different algebra than the usual fermionic one.  Certainly,  for $q=1$ we get $\hat{H_k}=\hat{n}_k\epsilon'_k$.
\section{Entropy} \label{S}
The corresponding entropies were defined in \cite{MMD1} and the fermion case was previously discussed in \cite{CMP} where  the number distributions in Eq. (\ref{n})
were obtained with use of  the maximum entropy principle. Here  we just want to find the thermodynamic expression for the entropy in terms of the average internal energy $<U>$ and the occupation number $<N>$.  We define 
\begin{equation}
S=\sum_k\Theta_k +\beta<U>- \beta\mu <N>,
\end{equation}
where the functions  $\Theta_k$ are to be determined. A simple calculation gives 
\begin {equation}
\beta<U>- \beta\mu <N>=\sum_k\frac{1}{q-1}<{N}_k>^{1/q}\left((1\mp <{N}_k>^{1/q})^{(q-1)}-<{N}_k>^{\frac{q-1}{q}}\right),
\end{equation}
where the upper sign applies to the fermionic case.  The entropy functions obtained in \cite{MMD1} and  \cite{CMP} 
 are reproduced if we define the
$\Theta_k$ functions as
\begin{equation}
\Theta_k=\mp\log_q\left(1\mp<{n}_k>^{1/q}\right),
\end{equation}
where $\log_q x=\frac{1-x^{q-1}}{1-q}$.  In addition, starting with the differential equation
\begin{eqnarray}
\frac{d<{N}_k>}{d\beta \epsilon_k'}&=&-q<{N}_k>^{(2q-1)/q}\left(1\mp <{N}_k>^{1/q}\right)^{2-q},\nonumber\\
                                                          &=& \Psi(<{N}_k>)
\end{eqnarray}
and  use of the Inverse Maximum Entropy Principle, \cite{KK},  the second derivative of the entropy with respect to $<{N}_k>$  
\begin{equation}
S''= \frac{c}{\Psi},
\end{equation}
leads to the entropy functions
\begin{equation}
S=\mp \left(1\mp <{N}_k>^{1/q}\right)\log_q\left(1\mp <{N}_k>^{1/q}\right)-<{N}_k>^{1/q} \log_q<{N}_k>^{1/q}.
\end{equation}
\section { Diffusion differential equations}\label{AD}
It is well known that one can obtain differential equations from the unnormalized density matrix \cite{RF}.  For the standard, $q=1$,  density matrix 
the corresponding  simplest differential equation is 
\begin{equation}
-\frac{\partial\rho_U}{\partial \beta}=H\rho_U,  \label{D}
\end{equation}
where $\rho_U=e^{-\beta H}$ is the unnormalized density matrix.  Letting $H=- \frac{\partial^2}{\partial x^2}$ will certainly leads to the standard diffusion equation. In our case, from Eq. (\ref{H})  we obtain that the corresponding operator is given by
\begin{equation}
{\rho}_U=\left(1+\beta (q-1){H}_k\right)^{\frac{1}{1-q}},
\end{equation}
leading to the nonlinear  differential equation
\begin{equation}
-\frac{\partial\rho_U}{\partial \beta}=H\rho_U^q. \label{dh}
\end{equation}
The simplest case is to consider $H=-\frac{\partial^2}{\partial x^2}$ giving  the differential equation
\begin{equation}
\frac{\partial \rho_U}{\partial \beta}=\frac{\partial^2}{\partial x^2}\rho_U^q,
\end{equation}
whose solution
\begin{equation}
\rho_U(x,t)\propto \frac{1}{\beta^{1/(1+q)}}\left(1-(q-1)\frac{x^2}{\beta^{2/(1+q)}}\right)^{1/(q-1)}.
\end{equation}
  was studied, among other applications, in the context of non-extensive Statistical Mechanics \cite{PP}, anomalous diffusion in the presence of external forces \cite{TB},  anomalous diffusion on fractals \cite{ MMPL}\cite{PMML}, and $\kappa$-generalized Statistical Mechanics \cite{WS}.

A more general differential equation can be obtained if we use the differential operator
\begin{equation}
 H=-\frac{\partial}{\partial x}\left(\frac{\partial f}{\partial x}\right)^{\mu}, \label{d}
\end{equation}
where in our case the function $f=\rho_U^q$ and $\mu$ is a parameter.
	This differential operator $\frac{\partial}{\partial x}\left(\frac{\partial }{\partial x}\right)^{\mu}$ was obtained, \cite{U1},  from the Fisher-like measure that results when one expands the relative entropy 
\begin{equation}
H_{\mu}(p(x)||p(x+\Delta)=\int   p(x) \left(-\ln \frac{p(x+\Delta)}{p(x)}\right)^{\mu} dx,\label{rl}
\end{equation}
up to second order with respect to a small shift $\Delta$.   The origin of this   relative entropy,\cite{U2},  is the entropic function
\begin{equation}
S_{\mu}(p)=\sum_ip_i(-\ln p_i)^{\mu},
\end{equation}
obtained by applying the Weyl fractional derivative to the function $\sum_i{ p_i^{-t}}$ and then taking the limit $t\rightarrow -1$.  As it is well known, the use of an ordinary derivative will
lead to the Shannon entropy, and the use  \cite{A}  of the  Jackson $q$-derivative \cite{J} will reproduce  the Tsallis entropy \cite{T}.  In addition, similar operators to the one   in Eq. (\ref{d}) have been studied in Refs. \cite{P1}-\cite{S} as an application to nonlinear diffusion  in the context of disturbances in a non-Newtonian fluid, non-linear heat conduction and fractal diffusion.  
Equations (\ref{dh}) and (\ref{d})  give the nonlinear differential equation
\begin{equation}
\frac{\partial \rho_U}{\partial \beta}=\frac{\partial}{\partial x}\left(\frac{\partial}{\partial x}\rho_U^q\right)^{\mu}. \label{eq}
\end{equation}
It is natural to consider the ansatz
\begin{eqnarray}
\rho_U&=&\frac{A}{\beta^{\lambda}}\Phi^{\omega}(x,\beta),\nonumber\\
\Phi(x,\beta)&=&\left(1+\frac{(1-q\mu)}{C}\frac{x^{\gamma}}{\beta^{\alpha}}\right). \label{Psi}
\end{eqnarray}
where the constants $\lambda$, $\gamma$ and $\alpha$  in the $q\rightarrow 1$ limit become: $\lambda=1/2$,  $\gamma=2$ and  $\alpha=1$. After  performing the elementary derivatives
and comparing powers in $\Phi(x,t)$, $x$ and $\beta$ and the constants in both sides of Eq. (\ref{eq}) we find 
\begin{eqnarray}
\omega&=&\frac{1}{\mu q-1},\\
\gamma&=&\frac{1+\mu}{\mu},\\
\alpha&=&\frac{1+\mu}{\mu^2(1+q)},\\
\lambda&=&\frac{1}{\mu(1+q)},\\
A^{1-q\mu}&=&\left(\frac{q(1+\mu)}{C\mu}\right)^{\mu}\mu(1+q),  \label{A}
\end{eqnarray}
where  the requirement that the constant $A$ has to be positive will restrict the values $\mu=\frac{n}{m}$,  where $n$ and $m$  are odd numbers.   The general solution is given by
\begin{equation}
\Phi(x,t)=\frac{A}{\beta^{\frac{1}{\mu(1+q)}}}\left(1+(1-\mu q)\frac{x^\frac{1+\mu}{\mu}}{\beta^{\frac{1+\mu}{\mu^2(1+q)}}}\right)^{-\frac{1}{1-\mu q}},
\end{equation}
leading for $q=1$ to the particular solution
\begin{equation}
\Phi(x,t)\sim \frac{1}{\beta^{1/2\mu}}\left(1+\frac{(1-\mu)x^{(1+\mu)/\mu}}{C\beta^{(1+\mu)/2\mu^2}}\right)^{-\frac{1}{1-\mu}},
\end{equation}
already obtained in \cite{U1} as a simple model for anomalous diffusion. An additional relation between the constants $A$ and $C$ can be obtained from the 
normalization of  the function in Eq. (\ref{Psi})
\begin{equation}
\frac{A}{\beta^{\lambda}}\int_{-\infty}^{\infty}\Phi^{\omega}(x,\beta) dx=1,
\end{equation}
which due to the fact that $\Phi(-x,\beta)=-\Phi(x,\beta)$ we can change  the limits from $(-\infty,\infty)$ to $(0,\infty)$ and with use of the integral representation of the $\Gamma(x)$ function we obtain
\begin{equation}
A=\frac{\gamma}{2C^{1/\gamma}}\frac{(1-q\mu)^{1/\gamma}\Gamma(-\omega)}{\Gamma(1/\gamma)\Gamma(-\omega-\frac{1}{\gamma})}.\label{A1}
\end{equation}
 In the particular case  of $q\mu=1$ there is an infinite number  of solutions involving the stretched exponential function
\begin{equation}
\Psi(x,\beta)=\frac{A}{\beta^{1/(\mu+1)}}exp\left(-\frac{x^{(1+\mu)/\mu}}{C\beta^{1/\mu}}\right),
\end{equation}
where from Eqs. (\ref{A}) and (\ref{A1}) the constants
\begin{equation}
A(q\mu=1)=\frac{1}{2\Gamma(\mu/(\mu+1))\mu^{(1-\mu)/(\mu+1)}},
\end{equation}
and
\begin{equation}
C(q\mu=1)=\frac{1}{\mu^2}(\mu+1)^{(\mu+1)/\mu},
\end{equation}
giving for the standard case ($q=1=\mu$)  the  expected values $A=\frac{1}{\sqrt{4\pi}}$ and $C=4$.
\section{Conclusions}
In this manuscript we obtained  for non-extensive statistical mechanics, according to the particle number distributions
discussed in Refs. \cite{CW}-\cite{CMP},   the corresponding density matrix and factorized partition function for the bosonic and fermionic cases
and a thermodynamic expression for the entropy functions. In addition, we showed that the unnormalized density matrix leads to a nonlinear
equation whose solution is a two-parameter function relevant to anomalous diffusion.  The use of the nonlinear differential operator in Eq.(\ref{eq}) gives an alternative
approach to study the diverse phenomena involving anomalous diffusion like for example  Levy flights, turbulent diffusion and two-dimensional rotating flow, other than the cases of the nonlinear 
Fokker-Planck equation \cite{F}, linear  \cite{MK}\cite{H}  and nonlinear \cite{BTG}  fractional Fokker-Planck equations.   A  continuation of the work done in this manuscript could include two 
different directions. One, would involve a calculation of the thermodynamic curvature, as done \cite{U3} for the case of the dilute gas approximation \cite{BDG} of non-extensive thermodynamics to learn about the stability and possible anyonic behavior of the system,  and as a matter of comparison with other approaches to anomalous diffusion phenomena, to extend the applications of Eq.(\ref{eq}) including a time dependent source term and external forces.
\subsubsection*{Acknowledgments}
I thank to the anonymous reviewers for their helpful suggestions.
 

\begin{thebibliography}{99}
\bibitem{T1} A list of works can be found in tsallis.cat.cbpf.br/biblio.htm.
\end{thebibliography}
\end{document}